# Energy-Efficient Data Collection in Clustered Wireless Sensor Networks employing Distributed DCT

Minh T. Nguyen *† and Keith A. Teague †

Thai Nguyen University of Technology, Vietnam
† Oklahoma State University, USA

**ABSTRACT**

*In this paper, a energy-efficient data collection method is proposed in which an integration between Discrete Cosine Transform (DCT) matrix and clustering in wireless sensor networks (WSNs) is exploited.Based on the fact that sensory data in WSNs is often highly correlated and is suitable to be transformed in DCT domain, we propose that each cluster from the networks only sends a small number of large DCT transformed coefficients to the base-station (BS) for data collection in two common ways, either directly or in multi-hop routing. All sensory data from the sensor network can be recovered based on the large coefficients received at the BS. We further analyze and formulate the communication cost as the power consumption for transmitting data in such networks based on stochastic problems. Some common clustering algorithms are applied and compared to verify the analysis and simulation results. Both noise and noiseless environments for the proposed method are considered.*

*Keywords*

*Wireless sensor networks, data compression, power consumption, discrete cosine transform*

## 1. INTRODUCTION

Saving energy in wireless sensor networks (WSNs) is a critical issue for many research studies.WSNs often consist of a large number of small inexpensive sensors with limited power,processing and computing resources [1], [2]. These sensors can sense, measure, and gather information from the environment and, based on some local decision processes, they can transmit the sensory data to the base-station (BS). Since the sensors often work in hash conditions without battery re-charge, they need to either reserve or save energy as much as they can to maintain their connections in such networks. There have been many research studies focusing on different network topologies, data processing and data collection methods to reduce power consumption for prolonging the network lifetime. We are expecting to have some energy-efficient combinations between those methods for further energy saving.

In WSNs, clustering algorithms have been shown to be energy efficient methods to collect data to the BS [3], [4], [5], [6], [7], [8]. There are different clustering algorithms that focus on differentparameters of the networks. The main goal is to balance and to reduce power consumption for all sensors. In [4], [5], some sensors are randomly chosen as cluster-heads (CH) and the rest choose the closest CHs to join to form clusters. Since the power consumption usually falls on CHs, sensors take turns to be CHs that can help balance energy for the entire network. Load balancing is studied in [6] in order to prolong network lifetime. In [7], the distance between CHs and non-CH sensors can be considered as a certain number of hops based on sensor transmission ranges. The total power consumption for the network is analyzed and minimized based on the hop distances. HEED [8] provides an algorithm to choose CHs based on the sensor residual energy that also helps sensors deplete energy equally. Two very common clustering methods, Kmeans [3] and LEACH [4], are chosen to compare with our analysis in this work that will be shown in the simulation results.

In order to reduce further power consumption for collecting data in WSNs, compressive sensing(CS) [9] is integrated in many different data routing methods [10], [11], [12], [13], [14], [15],[16], [17], especially in clustering methods [18], [19], [20], [21]. Based on the idea that sensory data can be represented in a small number of large transformed coefficients ($K$) in some proper domains, the network only needs to collect a certain number of CS measurements, denoted as $M$, to be able to recover all data from the sensing area that needs to be observed ($M > 2K$).

The number of measurements required are much smaller than the number of sensors ($N$) in such networks [22] ($M \ll N$).

It is motivated that if the sensors could send only a certain number of the large coefficients ($K$)for data collection and the BS can recover all data based on the received coefficients. Discrete Cosine Transform (DCT) have been utilizing efficiently for numerous applications, especially in data processing. Papers [23], [24] study data compression for WSNs applying DCT and Wavelet, respectively. An algorithm called RIDA [25] proposed a novel paradigm to compress data in such networks using logical mapping. Sensor readings in each cluster are sorted within each cluster.Virtual indices are assigned to sensors based on their orders and are sent to the BS for a mapping process. Each sensor calculates its own data with DCT coefficients. Only the large products are sent to the BS. None of the algorithms mentioned above consider the power consumption for communication in such networks.

In this paper, a distributed algorithm is proposed to transmit only the large DCT transformed coefficients to the BS for the signal recovery processes. A WSN is partitioned into non-overlapped clusters by two common clustering algorithms Kmeans [3] and LEACH [4]. At the beginning, all sensors send their raw data to their own CHs which they belong to. These CHs sort the received data and multiply to a DCT sub-matrix to achieve a certain number of large coefficients, and finally send these coefficients to the BS. We offer and also analyze two ways to forward the coefficients from CHs to the BS, directly and in multi-hop routing. At the BS, in order to recover the raw data for a cluster, a transformed vector is built by using the large coefficients received from the corresponding CH. The rest of the vector is chosen to be zeros. We analyze and formulate the power consumption for transmitting all data in such networks and provide simulation results to be compared with the analysis. The compression algorithm is simulated and addressed with real sensor readings in both noise and noiseless environments.

In summary, the contributions of our paper are

1) A new distributed DCT data compression for data collection in WSNs is proposed.
2) All transmission power consumptions for the network are formulated, analyzed and simulated.
3) The proposed method is considered with both noiseless and noisy environments. Some optimal cases for the networks are suggested.
The rest of this paper is organized as follows. The main problems and the algorithm are addressed in Section II. The power consumption for all data transmissions is calculated in Section III. Simulation results are shown in Section IV. Finally, conclusions and future work are presented in Section V.

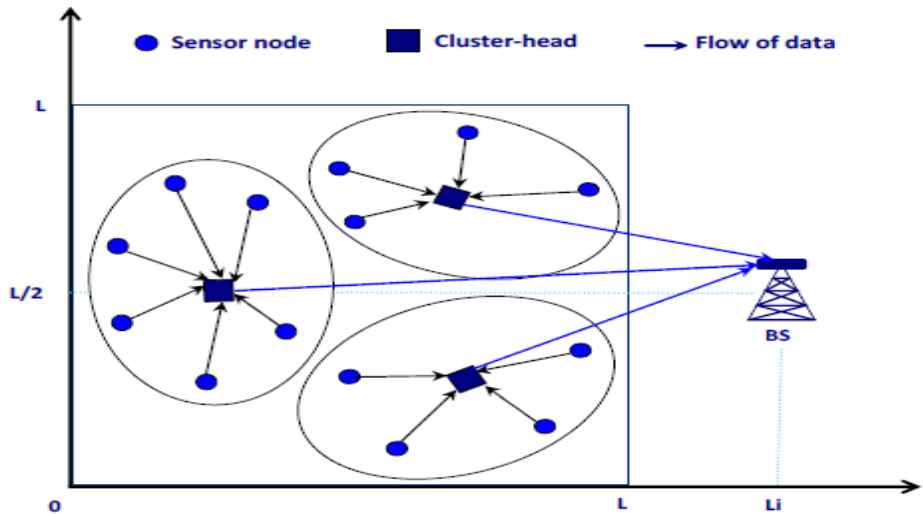

Fig. 1. A clustered WSN with BS outside the sensing area ($Li > L$).

## 2. PROBLEM FORMULATION

*A. Network model*

In our network model, we assume that N sensors have been distributed randomly with equal probability in either a square sensing area (dimension $L \times L$) or a circular area with the radius $R0$. The network is partitioned into *Nc* clusters. The CHs are randomly chosen from all the sensors based on a probability *Nc/N*. The non-CH sensors choose one CH which is closest to form clusters. In our analysis, we assume all clusters have an equal number of sensors. On average, each cluster has $(\frac{N}{N_c} - 1)$ non-CH sensors and one CH which is also randomly distributed in the sensing area. The BS can be outside or inside the sensing area following this fixed positio $(L_i, \frac{L}{2})$ as shown in Figure 1.

*B. DCT data compression algorithm*

Figure 1 shows the clustered network in general with the BS outside the sensing area. As mentioned in the previous section, we need to observe all sensory readings from the network and send them to the BS. DCT is chosen to create a square sub-matrix $\phi_i$ with dimension ($n_i \times n_i$) at each CH, where *ni* is the number of sensors in the $i^{th}$ cluster. All readings from non-CH sensors are sent to CHs with their indices and then are sorted in descending or ascending order. The significant differences between sorted and unsorted data are presented in the simulation section. After being multiplied with a sparsifying DCT matrix, a large proportion of the signal energy is focused on the very first large coefficients, given as *K* coefficients. Only these coefficients all CHs are sent to the BS. The rest of the transformed vector can be considered as zeros which are added back to the large coefficients at the BS for the recovery process. By sending only *K* values to the BS, this method can save significant energy for data collection. We discuss the trade-off later in the simulation section. The DCT compression algorithm can be written in short as:

1) Non-CH sensors send their own readings to their CHs.
2) All received data including the CHs' reading are sorted and then are transformed in DCT domain.
3) There are only *K* large coefficients taken to be sent to the BS. The rest of the transformed coefficients are considered as zeros.
4) At the BS, all readings from each cluster are recovered by multiplying the large coefficients to the sub-matrix used at each cluster.

In formula, we have the transformed vector $s_i$ at the cluster $i^{th}$ as follows

$$\underline{s_i} = \phi_i \underline{x_i}. \qquad (1)$$

At the BS, we obtain the readings from the $i^{th}$ cluster based on $\underline{s'_i}$ which is created from the received *ki* large coefficients and additional zeros as

$$\underline{\hat{x}_i} = \phi_i^T \underline{s'_i}. \qquad (2)$$

Equation (3) shows coefficients of the DCT matrix in general, which is returned by the function *dctmtx* in Matlab.

$$\Phi_{p,q} = \begin{cases} \frac{1}{\sqrt{N}}, & p = 0; \ 1 \leq q \leq N-1 \\ \sqrt{\frac{2}{N}} \cos(\frac{\pi(2q+1)p}{2N}), & 1 \leq (p,q) \leq N-1 \end{cases} \quad (3)$$

This algorithm can also work well with fault tolerance in the network since all sensors take turns to become CHs. Each node is only responsible to the others within its cluster. Network or node faults could be detected and recovered by fault tolerant algorithms for clustered networks [26] or for tree-based network [27] since we apply multi-hop routing to relay the coefficients. In that case, malfunctioned nodes are isolated but could be used for relaying data in the network if possible. This could be an open work for our future research.

## 3. POWER CONSUMPTION ANALYSIS

The total power consumption for transmitting and receiving data in WSNs [28], denoted as $P_{Tx}$ and $P_{Rx}$, are usually calculated, respectively as

$P_{Tx} = P_{T0} + PA(d)$ (4)

And

$P_{Rx} = P_{R0}$, (5)

where $P_{T0}$ and $P_{R0}$ are electronics consuming power depending on some elements such as coding, modulation, and signal processing. These factors do not depend on transmitting distances, denoted as $d$. Only the consumed power of the power amplifier $PA(d)$ is a function of $d$ which we consider to formulate based on stochastic problem in this paper. The total power consumption for transmitting data in such networks contain two parts, the intra-cluster power consumption denoted as $P_{intra-cluster}$ is for non-CH sensors to transmit their readings to the CHs and the consumed power for all CHs to transmit the large transformed coefficients to the BS, denoted as *PtoBS*.

### A. Analysis of $P_{intra-cluster}$

We assume to have a uniformly distributed WSN divided into *Nc* non-overlapped clusters with the same number of sensors as *N/Nc*, consisting of one CH and $(\frac{N}{N_c} - 1)$ non-CH nodes. We have

$$P_{intra-cluster} = N_C(\frac{N}{N_c} - 1) E[r^\alpha], \quad (6)$$

where $r$ is a random variable which represents distances between non-CH sensors to CHs they belong to, and $\alpha$ is the path loss exponent. As mentioned in [29], $\alpha = 2$ *or* 4 in free space or multiple fading channels, respectively. For simplicity, we assume it to be 2 throughout the paper. We can calculate $E[r^2]$ as follows:

$$E[r^2] = \int\int (x^2 + y^2)\, \rho(x,y)\, dx\, dy \tag{7}$$

$$= \int\int r'^2 \rho(r', \theta)\, r'\, dr'\, d\theta. \tag{8}$$

in which $\rho(x, y)$ is the node distribution. Similar to [4], we assume that each cluster area is circular with radius $R = L/\sqrt{\pi N_c}$ and the density of the nodes is uniform throughout the cluster area, i.e. $\rho(r', \theta) = 1/(L2/Nc)$. Equation (8) can be shown as

$$E[r^2] = \frac{1}{(L^2/N_c)} \int_{\theta=0}^{2\pi} \int_{r'=0}^{R} r'^3 dr'\, d\theta, \tag{9}$$

and finally we obtain

$$E[r^2] = \frac{L^2}{2\pi N_c}, \tag{10}$$

and the total intra-cluster power consumption

$$P_{intra-cluster} = (\frac{N}{N_c} - 1)\frac{L^2}{2\pi} \tag{11}$$

We can see that the total intra-cluster power consumption is a decreasing function of the number of clusters.

*Note*: in case the sensing are is circular, $P_{intra-cluster}$ is calculated as follows

$$P_{intra-cluster} = (\frac{N}{N_c} - 1)\frac{R_0^2}{2}, \tag{12}$$

where $R_0$ is the radius of the sensing area.

International Journal of Wireless & Mobile Networks (IJWMN) Vol. 8, No. 5, October 2016

*B. Analysis of $P_{toBS}$*

We assume that all clusters have the same number of sensors. We show that the number of large coefficients taken from a cluster are linearly proportional to the number of sensors in that cluster. Hence, in our analysis case, the number of large coefficients collected from *Nc* clusters should be equal. The total number of large coefficients is calculated as

$$K = \sum_{i=1}^{N_c} k_i = N_c k_i, \tag{13}$$

where *ki* is the number of coefficients collected from cluster *ith*. We consider both ways to forward the large coefficients to the BS, transmitting directly or forwarding through intermediate CHs based on a routing tree.

*1) Transmitting the large coefficients directly to the BS:* As shown in Figure 1, the average consumed power for all CHs to transmit *K* large coefficients to the BS is

$$P_{toBS} = KE[d^2], \tag{14}$$

where *d* is the random variable representing the distance between CHs and BS. Assuming that all CHs are randomly distributed in the entire area to balance the power consumption for the network, the expected squared distance between CHs and the BS [15] is given by

$$E[d^2] = \int_0^L \int_0^L [(x - L_i)^2 + (y - \frac{L}{2})^2] f(x,y) dx dy \tag{15}$$

$$= \frac{1}{L}[\frac{(L - L_i)^3}{3} + \frac{L_i^3}{3}] + \frac{L^2}{12}, \tag{16}$$

where $f(x,y) = \frac{1}{L^2}$ is uniform distribution of CHs in the sensing area.

From Equations (11) and (16), the total power consumption in this method can be formulated in general as

$$P_{total} = (\frac{N}{N_c} - 1)\frac{L^2}{2\pi} + \frac{K}{L}[\frac{(L - L_i)^3 + L_i^3}{3}] + \frac{KL^2}{12}. \tag{17}$$

When the BS is at the center of the sensing area (*Li = L/2*), Equation (16) is simplified as $E[d^2] = \frac{L^2}{6}$, and the total power consumption for the network is

$$P_{total} = (\frac{N}{N_c} - 1)\frac{L^2}{2\pi} + \frac{KL^2}{6}. \qquad (18)$$

*Note*: if the sensing area is circular and the BS is at the center, *PtoBS* is calculated in [20] as

$$E[d^2_{toBS}] = \frac{R_0^2}{2}. \qquad (19)$$

Hence, the total power consumption is

$$P_{total} = (\frac{N}{N_c} - 1)\frac{R_0^2}{2} + K\frac{R_0^2}{2}. \qquad (20)$$

*2) Transmitting the large coefficients to the BS using inter-cluster multi-hop routing:*

As shown in Figure 2, all sensors are randomly deployed in a circular sensing area with the BS at the center. Since CHs are randomly chosen from the sensors based on a probability *Nc/N*, they are also randomly distributed in the area. We assume to have an algorithm to form a routing tree connecting all the CHs with the root as the BS at the center of the sensing area as mentioned in [20]. The distance between a random CH and the BS can be considered as a random variable, denoted as *x*. The probability of being able to make a connection at distance *x* using *hops* or less hops is denoted by $P_{hops(x)}$. In paper [30] the mean value of the number of hops (*hops*) is calculated as follows

$$E[hops] = max(hops) - \sum_{n=1}^{max(hops)-1} \frac{P_{hops}(x)}{P_{max(hops)}(x)}, \qquad (21)$$

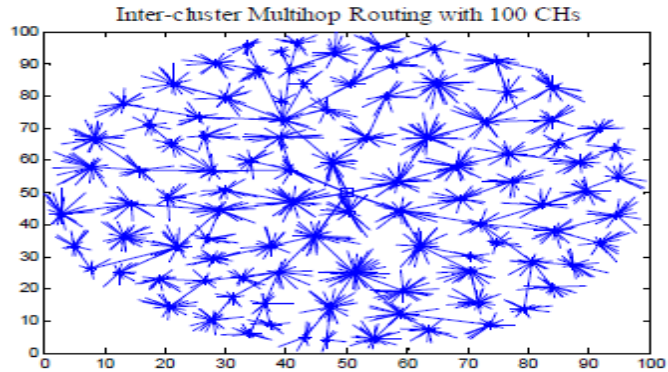

Fig. 2. Transmissions in the network with inter-cluster multi-hop routing when the BS at the center

International Journal of Wireless & Mobile Networks (IJWMN) Vol. 8, No. 5, October 2016

where *max(hops)* is the maximum number of hops allowed. Finally, we obtain the total consumed power for relaying *K* large coefficients from CHs to the BS as

$$P_{toBS} = \left\{ hops_{max} - \sum_{hops=1}^{hops_{max}-1} \frac{P_{hops}(x)}{P_{max(hops)}(x)} \right\} R^2 K, \qquad (22)$$

where *R* is the CH's transmission range. This value can be changed depending on the CH density. In other words, if the number of CHs reduces, we need to increase *R* to maintain all CHs connected as a routing tree. We have the total power consumption for data collection in the entire network using multi-hop relaying as

$$P_{total} = (\frac{N}{N_c} - 1)\frac{L^2}{2\pi} + E[hops]\, R^2 K. \qquad (23)$$

From Equations (18), (20) and (23), the total communication power consumptions are the linear functions of the number of large coefficients *K*.

## 4. SIMULATION RESULTS

In this section, we consider both types of networks, square sensing area with dimension 100 × 100 and circular area with radius $R_0 = 50$. 2000 sensors are randomly distributed in the areas. K-means [3] and LEACH [4] are applied as two common clustering algorithms to be compared with our analysis; K-means minimizes the intra-cluster power consumption while LEACH balances the power depletion between CHs and non-CH sensors. We consider both sorted and unsorted signals collected from Sensorscope: Sensor Networks for Environmental Monitoring [31]. These types of data provide different values of *K* that affect not only the transmitting cost from the CHs to the BS but also the reconstruction error at the BS. We first show the power consumption for the network and then the results of DCT compression. For the reconstruction error related to signal recovery, the normalized reconstruction error $\frac{\|x-\hat{x}\|_2}{\|x\|_2}$ is applied.

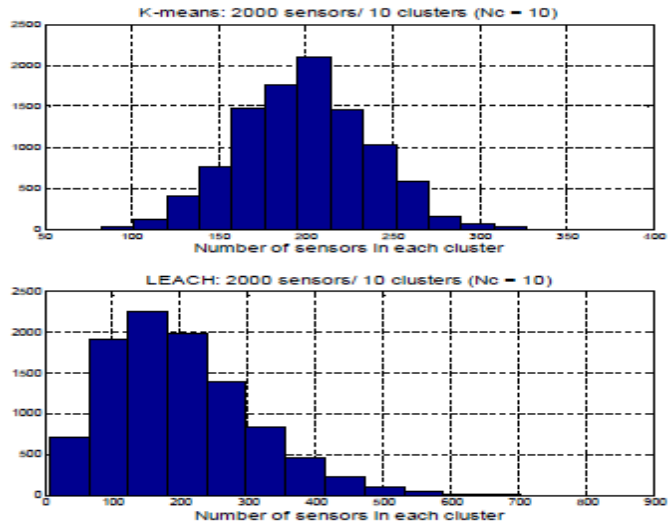

Fig. 3. Histograms from two clustering algorithm Kmeans and LEACH in the network with 2000 sensors distributed in asquare area 100 *x* 100

Figure 3 compares the histogram of the number of sensors in each cluster between two clustering algorithms, Kmeans and LEACH. We can see that Kmeans provides more uniform size for clusters than LEACH does, which results in the smaller intra-cluster power consumption as shown in Figure 4. It is shown that the intra-cluster consumed power reduces as the clusters become smaller. The total consumed power for the network to transmit data to the BS at the position of $L_i = 3L$

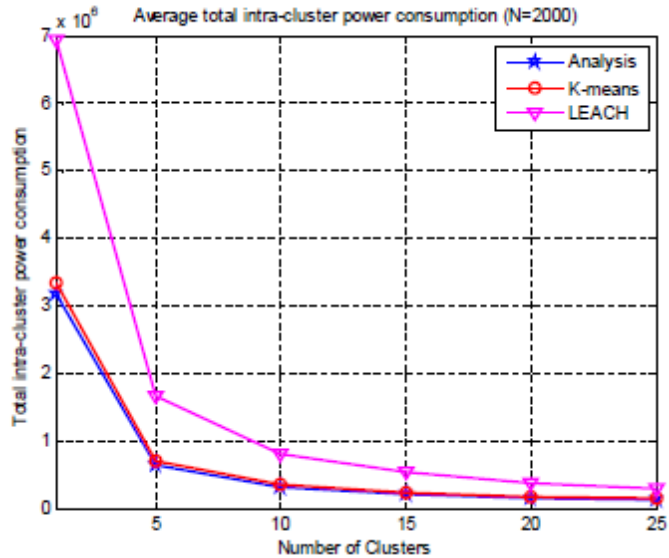

Fig. 4. Total intra-cluster power consumption for the network with 2000 sensors distributed in a square area 100 *x*100

International Journal of Wireless & Mobile Networks (IJWMN) Vol. 8, No. 5, October 2016

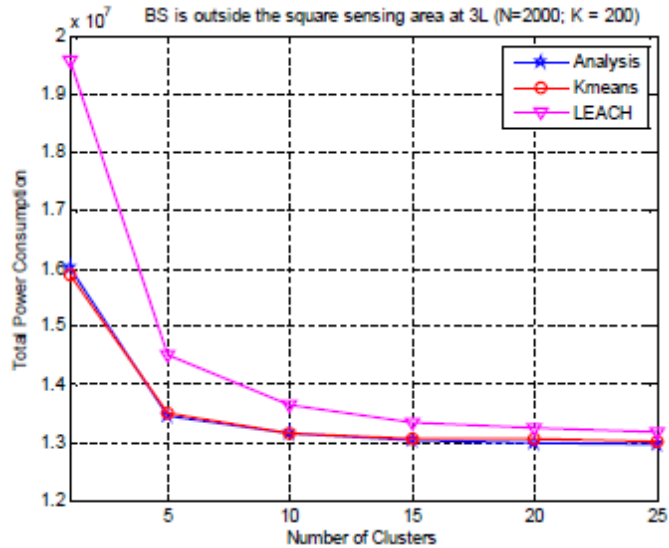

Fig. 5. Total power consumption for all transmissions in the network with 2000 sensors partitioned into different number of clusters when the BS is outside the sensing area at 3L

is shown in Figure 5. Since the total number of large coefficients is chosen as $K = 200$, this power is decreased as we increase the number of clusters following the intra-cluster power consumption.

In order to work in the circular sensing area with multi-hop routing, we use a greedy algorithm proposed in [20] to form a routing tree between CHs. At each network divided into different numbers of clusters, we use different transmission ranges $R = [50\ 30\ 25\ 22\ 18]$ corresponding

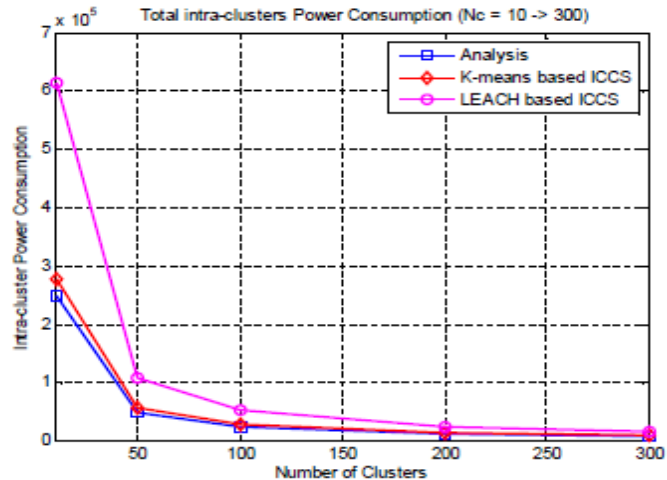

Fig. 6. Total intra-cluster power consumption for the network with 2000 sensors distributed in a circular sensing area with radius $R_0 = 50$

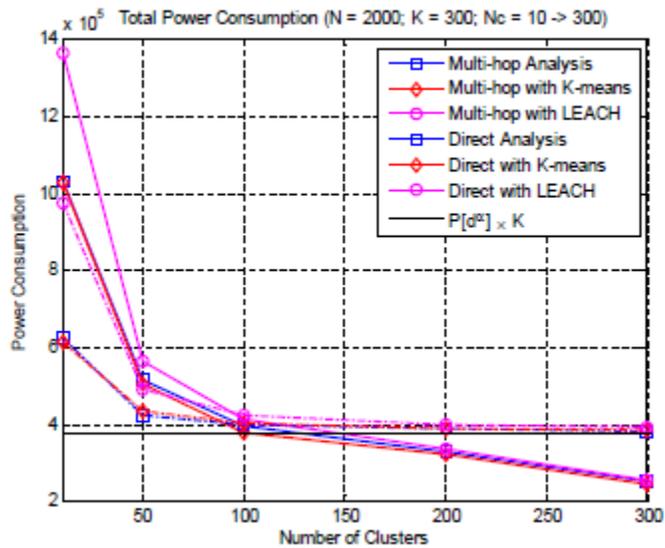

Fig. 7. Total power consumption for all transmissions in the network using inter-cluster multi-hop to forward K large coefficients the the BS in a circular sensing area with radius $R_0 = 50$

to $N_c$ = [10 50 100 200 300]. Figure 6 show the intra-cluster power consumptions calculated from K-means, LEACH and the analysis case. Figure 7 shows and compares the total power consumptions for both methods, direct and multi-hop to forward the coefficients to the BS. It is shown that, if the numbers of clusters are small, the direct method still consumes less power than the multi-hop method. As the number of clusters is greater than 150, the multi-hop routing methods consumes less power than the direct one.

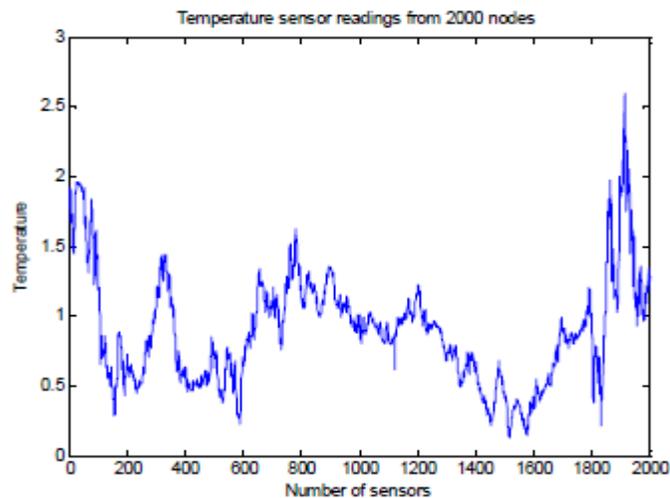

Fig. 8. Unsorted sensory readings collected from 2000 sensors

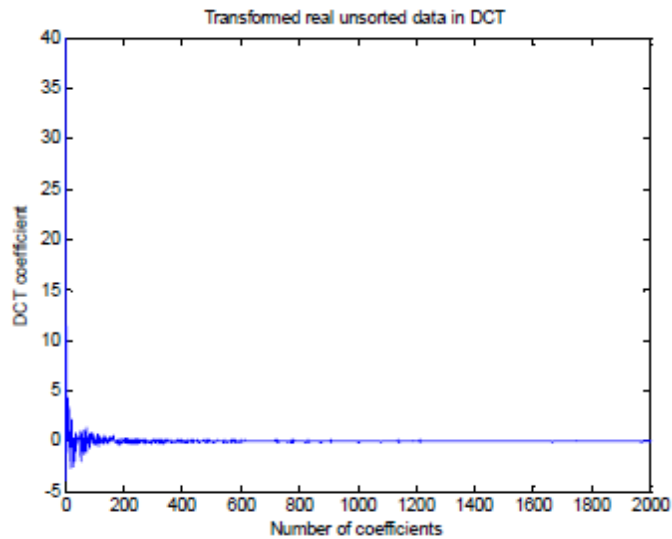

Fig. 9. DCT transformed coefficients from 2000 unsorted sensory readings

Figure 8 shows unsorted sensor readings and Figure 9 shows their transformed coefficients in the DCT domain. All signal energy is preserved in the transformed vector but is now focused in relatively small numbers of large coefficients. If we transmit only these $K$ large valued to the BS, this results in much less consumed power than transmitting all the readings.

Figure 10 shows sorted signals in decreasing order. The DCT transformed coefficients are shown in Figure 11. In this case, the large coefficients are concentrated in the lower numbered coefficients. The transmission cost can be reduced based on the smaller values of $K$ compared to that in unsorted signals shown in Figure 9. That is the reason we chose to sort sensing data at each cluster.

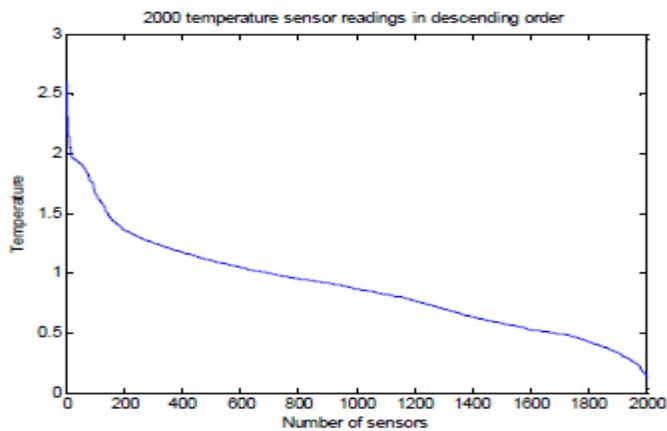

Fig. 10. Descending sorted sensory readings collected from 2000 sensors

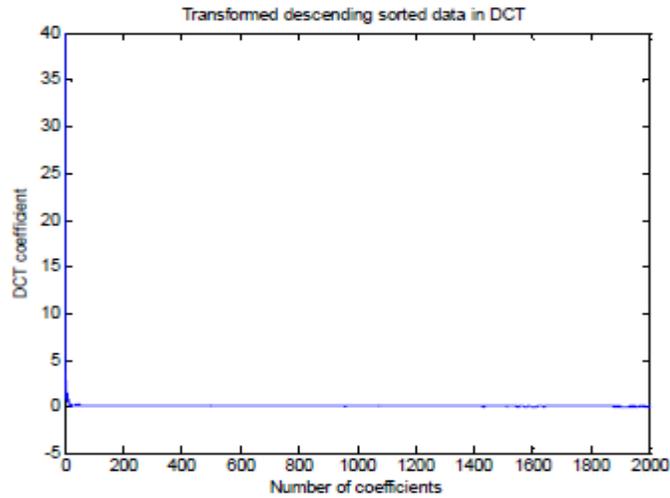

Fig. 11. DCT transformed coefficients from 2000 sorted readings

Figure 12 shows that both types of sorted data result in the same reconstruction error in different numbers of clusters. The large coefficients in sorted data are focused in the smaller numbered coefficients. With the same number of coefficients being sent to the BS, the ones from sorted signals carry more signal energy than unsorted signals do. That explains the smaller reconstruction errors with sorted data.

As shown in Figure 13, increasing the number of clusters or reducing the total number of coefficients *K* transmitted to the BS results in increasing the reconstruction errors. Transmitting more of the larger DCT coefficients to the BS can compensate for the errors as we increase the number of clusters.

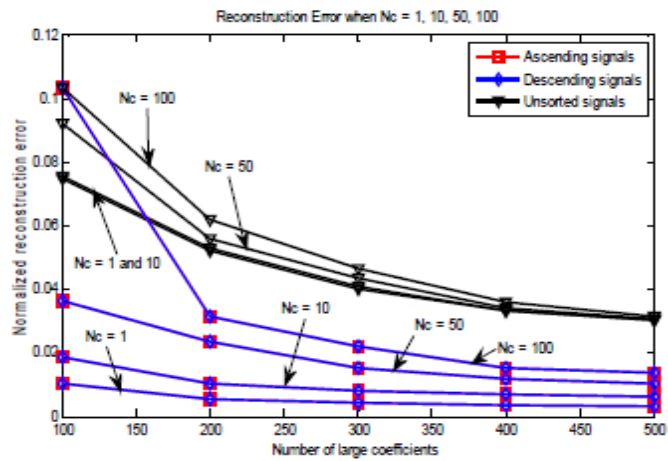

Fig. 12. Reconstruction error versus number of large coefficients with different number of clusters

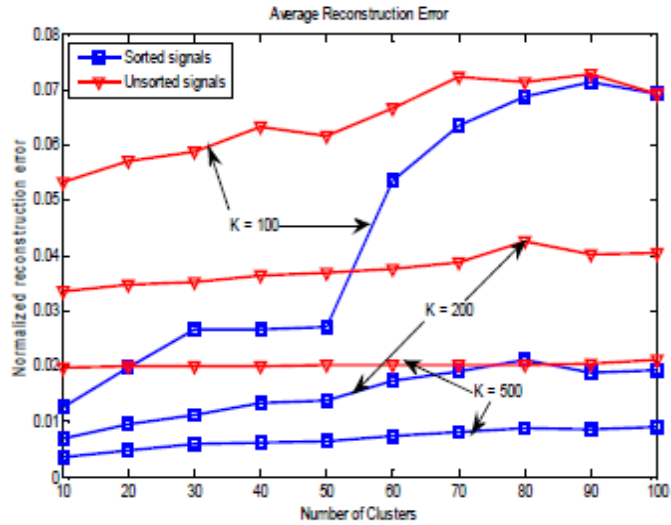

Fig. 13. Reconstruction error versus number of clusters with different number of the large coefficients (K)

In a noiseless environment, using DCT compression consumes very little power since the network only sends *K* large transformed coefficients ($K \ll N$). As shown in our simulation results, *K* is generally only about 10% as large as *N* to satisfy an error-target in signal recovery processes. In practical networks, noise is problematic. DCT compression is quickly degraded as shown in Figure 14. The reconstruction errors keep increasing as the total number of large coefficients increase.

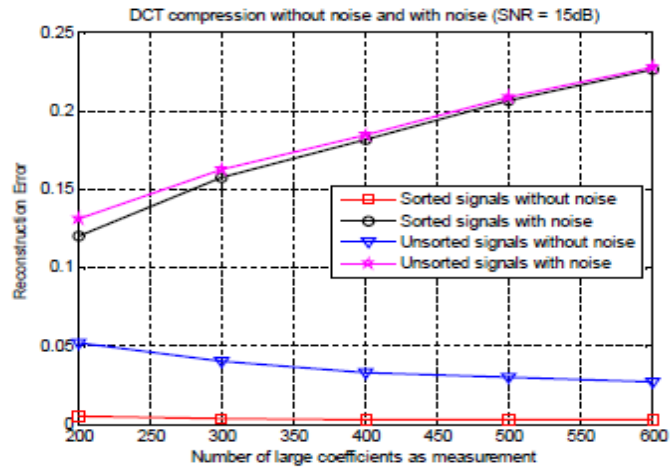

Fig. 14. DCT compression reconstruction error versus the number of large coefficients with noise and noiseless

## 5. CONCLUSIONS AND FUTURE WORK

In this paper, we proposed DCT based data compression algorithms for clustered WSNs to reduce power consumption in collecting sensory data. Based on the fact that almost all data energy focused in relatively small numbers of large coefficients in the transformed vectors, we only send the large coefficients to the BS for the signal recovery process. These coefficients are mapped at the BS to recover all sensory readings from the network. The proposed method significantly minimize the amount of data transmission in such networks. We analyzed and formulated either the intra-cluster power consumption or the total power consumption for the network to transmit data. Simulation results are provided for both consumed power calculation and the DCT compression method. We concluded that this DCT compression method degrades its performance when working in noisy environment. We suggest an optimal case for the networks to use multi-hop if many clusters are applied. In future work, we will study the boundary for the number of clusters in both noise and noiseless environments.

## 6. ACKNOWLEDGEMENTS

This work is supported by Thai Nguyen University of Technology (TNUT), Vietnam and School of Electrical and Computer Engineering (ECE), Oklahoma State University, USA.

## REFERENCES

[1]  J. Yick, B. Mukherjee, and D. Ghosal, "Wireless sensor network survey," Computer Networks, vol. 52, no. 12, pp. 2292– 2330, 2008.
[2]  I. Akyildiz, W. Su, Y. Sankarasubramaniam, and E. Cayirci, "Wireless sensor networks: a survey," Computer Networks,vol. 38, no. 4, pp. 393 – 422, 2002.
[3]  J. B. MacQueen, "Some methods for classification and analysis of multivariate observations," in Proc. of the fifth Berkeley Symposium on Mathematical Statistics and Probability (L. M. L. Cam and J. Neyman, eds.), vol. 1, pp. 281–297, University of California Press, 1967.
[4]  W. Heinzelman, A. Chandrakasan, and H. Balakrishnan, "An application-specific protocol architecture for wireless microsensor networks," Wireless Communications, IEEE Transactions on, vol. 1, pp. 660 – 670, Oct 2002.
[5]  M. Handy, M. Haase, and D. Timmermann, "Low energy adaptive clustering hierarchy with deterministic cluster-head selection," in Mobile and Wireless Communications Network, 2002. 4th International Workshop on, pp. 368 – 372, 2002.
[6]  G. Gupta and M. Younis, "Load-balanced clustering of wireless sensor networks," in Communications, 2003. ICC '03.IEEE International Conference on, vol. 3, pp. 1848 – 1852 vol.3, May 2003.
[7]  S. Bandyopadhyay and E. Coyle, "An energy efficient hierarchical clustering algorithm for wireless sensor networks," in INFOCOM 2003. Twenty-Second Annual Joint Conference of the IEEE Computer and Communications. IEEE Societies,vol. 3, pp. 1713 – 1723 vol.3, March-3 April 2003.

[8] O. Younis and S. Fahmy, "Distributed clustering in ad-hoc sensor networks: a hybrid, energy-efficient approach," in INFOCOM 2004. Twenty-third AnnualJoint Conference of the IEEE Computer and Communications Societies, vol. 1, pp. 4 vol. (xxxv+2866), March 2004.

[9] D.L.Donoho, "Compressed sensing," Information Theory, IEEE Transactions on, vol. 52, pp. 1289 – 1306, 2006.

[10] M. Rabbat, J. Haupt, A. Singh, and R. Nowak, "Decentralized compression and predistribution via randomized gossiping," in Information Processing in Sensor Networks, 2006. IPSN 2006. The Fifth International Conference on, pp. 51–59, 2006.

[11] M. T. Nguyen and K. A. Teague, "Compressive sensing based energy-efficient random routing in wireless sensor networks," in 2014 International Conference on Advanced Technologies for Communications (ATC 2014), pp. 187–192, Oct 2014.

[12] C. Luo, F. Wu, J. Sun, and C. W. Chen, "Efficient measurement generation and pervasive sparsity for compressive data gathering," Wireless Communications, IEEE Transactions on, vol. 9, pp. 3728–3738, December 2010.

[13] M. Nguyen and Q. Cheng, "Efficient data routing for fusion in wireless sensor networks," in The 25th International Conference on Computer Applications in Industry and Engineering (CAINE), New Orleans, LA, 2012, Nov 2012.

[14] J. Wang, S. Tang, B. Yin, and X.-Y. Li, "Data gathering in wireless sensor networks through intelligent compressive sensing," in INFOCOM, 2012 Proceedings IEEE, pp. 603–611, March 2012.

[15] M. T. Nguyen, "Minimizing energy consumption in random walk routing for wireless sensor networks utilizing compressed sensing," in System of Systems Engineering (SoSE), 2013 8th International Conference on, pp. 297–301, June 2013.

[16] M. T. Nguyen and K. Teague, "Tree-based energy-efficient data gathering in wireless sensor networks deploying compressive sensing," in Wireless and Optical Communication Conference (WOCC), 2014 23rd, pp. 1–6, May 2014.

[17] M. T. Nguyen and K. A. Teague, "Neighborhood based data collection in wireless sensor networks employing compressive 18 sensing," in 2014 International Conference on Advanced Technologies for Communications (ATC 2014), pp. 198–203,Oct 2014.

[18] M. T. Nguyen and N. Rahnavard, "Cluster-based energy-efficient data collection in wireless sensor networks utilizing compressive sensing," in Military Communications Conference, MILCOM 2013 - 2013 IEEE, pp. 1708–1713, Nov 2013.

[19] R. Xie and X. Jia, "Transmission-efficient clustering method for wireless sensor networks using compressive sensing," IEEE Transactions on Parallel and Distributed Systems, vol. 25, pp. 806–815, March 2014.

[20] M. T. Nguyen, K. Teague, and N. Rahnavard, "Inter-cluster multi-hop routing in wireless sensor networks employing mcompressive sensing," in Military Communications Conference (MILCOM), 2014 IEEE, pp. 1133–1138, Oct 2014.

[21] M. T. Nguyen and K. A. Teague, "Compressive sensing based data gathering in clustered wireless sensor networks," in 2014 IEEE International Conference on Distributed Computing in Sensor Systems, pp. 187–192, May 2014.

[22] J. Haupt, W. Bajwa, M. Rabbat, and R. Nowak, "Compressed sensing for networked data," Signal Processing Magazine, IEEE, vol. 25, pp. 92–101, March 2008.

[23] L. Makkaoui, V. Lecuire, and J. Moureaux, "Fast zonal dct-based image compression for wireless camera sensor networks," in Image Processing Theory Tools and Applications (IPTA), 2010 2nd International Conference on, pp. 126–129, July 2010.

[24] A. Ciancio and A. Ortega, "A distributed wavelet compression algorithm for wireless multihop sensor networks using lifting," in Acoustics, Speech, and Signal Processing, 2005. Proceedings. (ICASSP '05). IEEE International Conference on, vol. 4, pp. iv/825–iv/828 Vol. 4, March 2005.

[25] T. Dang, N. Bulusu, and W.-C. Feng, "Rida: A robust information-driven data compression architecture for irregular wireless sensor networks," in Proceedings of the 4th European Conference on Wireless Sensor Networks, EWSN'07,(Berlin, Heidelberg), pp. 133–149, Springer-Verlag, 2007.
[26] G. Gupta and M. Younis, "Fault-tolerant clustering of wireless sensor networks.," in WCNC, pp. 1579–1584, IEEE, 2003.
[27] L. Chitnis, A. Dobra, and S. Ranka, "Analyzing the techniques that improve fault tolerance of aggregation trees in sensor networks," J. Parallel Distrib. Comput., vol. 69, pp. 950–960, Dec. 2009.
[28] Q. Wang, M. Hempstead, and W. Yang, "A realistic power consumption model for wireless sensor network devices," in Sensor and Ad Hoc Communications and Networks, 2006. SECON '06. 2006 3rd Annual IEEE Communications Society on, vol. 1, pp. 286–295, Sept 2006.
[29] T. S. Rappaport, Wireless Communications: Principles and Practice (2nd Edition). Prentice Hall, 2 ed., Jan. 2002.
[30] S. Chandler, "Calculation of number of relay hops required in randomly located radio network," Electronics Letters, vol. 25, no. 24, pp. 1669–1671, 1989.
[31] http://lcav.epfl.ch/op/edit/sensorscope en.